\documentclass[12pt]{iopart}
\usepackage{psfig,graphicx}

\newcommand{\diag}{{\rm diag\,}}
\newcommand{\Det}{{\rm Det\,}}

\newcommand{\re}{{\rm Re\,}}
\newcommand{\im}{{\rm Im\,}}

\begin{document}

\title[A Supersymmetry Approach to Billiards with Randomly
       Distributed Scatterers]
      {A Supersymmetry Approach to Billiards with Randomly
       Distributed Scatterers II: Correlations}

\author{Thomas Guhr\dag \
        and Hans--J\"urgen St\"ockmann\ddag
\address{\dag\
         Matematisk Fysik, LTH, Lunds Universitet,
         Box 118, 22100 Lund, Sweden}
\address{\ddag\
         Fachbereich Physik der Philipps--Universit\"at Marburg,
         35032 Marburg, Germany}
        }

\begin{abstract}
In a previous contribution (H.J.~St\"ockmann, J. Phys. {\bf A35}, 5165
(2002)), the density of states was calculated for a billiard with
randomly distributed delta--like scatterers, doubly averaged over the
positions of the impurities and the billiard shape. This result is now
extended to the $k$--point correlation function.  Using supersymmetric
methods, we show that the correlations in the bulk are always
identical to those of the Gaussian Unitary Ensemble (GUE) of random
matrices. In passing from the band centre to the tail states, The
density of states is depleted considerably and the two--point
correlation function shows a gradual change from the GUE behaviour to
that found for completely uncorrelated eigenvalues. This can be viewed
as similar to a mobility edge.
\end{abstract}


\submitto{\JPA}


\section{Introduction}
\label{sec1}

The theory of random matrices provides a schematic, but powerful
statistical model for a wide class of spectral problems in complex
systems, for reviews see Refs.~\cite{Mehta,Haake,GMGW}. In particular,
there is overwhelming evidence for the fact that the spectral
fluctuations of a quantum system whose classical counterpart is fully
chaotic are described by the Gaussian ensemble of random matrices,
i.e.~by the Gaussian Unitary Ensemble (GUE) in the absence of time
reversal invariance and by the Gaussian Orthogonal Ensemble (GOE) if
time reversal invariance holds and the spectrum is free of Kramers
degeneracies~\cite{BGS}. On the other hand, the fluctuation properties
for quantum systems whose classical counterparts are regular ought
to be different, and often of the Poisson type. Many systems show
mixed fluctuation properties and transitions from regular to chaotic
behaviour.

Quantum billiards are ideal systems for the study of spectral
fluctuation properties. Billiards are said to be ballistic because the
classical dynamics and the quantum spectra are exclusively determined
by the shape of the boundary.  Whereas such ballistic systems are well
understood, the situation is less clear for disordered systems. In
particular, there are many open questions concerning the
localization--delocalization transition if disorder is varied. From
the one--parameter scaling hypothesis~\cite{Abr79} is is generally
accepted that in one-- and two--dimensional systems all states are
localized, but analytic proofs exist only for one--dimensional systems
(see reference \cite{Kra93} for a review). There are a number of work
using supersymmetric techniques, where the change of the wave function
amplitude statistic is studied with the reciprocal conductance as a
perturbation parameter~\cite{mir00}, but up to now there is no closed
theory covering the full range from localized to delocalized wave
functions. On the other hand there are microwave experiments showing a
clear localization--delocalization transition with frequency
\cite{Kud95,Stoe01b}.

This was the motivation of a previous publication \cite{St1}, hitherto
denoted Ref.~I, to tackle the problem by an alternative
approach. Instead of the usually applied non--linear $\sigma$--model
the more explicit system of a billiard with randomly distributed
scatterers was studied. This approach generalized a model introduced
by Bogomolny et al.~\cite{Bogo}. The average over disorder was
achieved with help of a trick using the conjecture that a typical
wavefunction can be viewed as a random superposition of plane
waves~\cite{ber77a}. Thereby, no supersymmetric field variables are
needed which Efetov used to construct his non--linear
$\sigma$--model~\cite{efe83}. It avoids as well the complications of
diagrammatic expansions of Green functions and summations of ladder
diagrams~\cite{Kra93}. It was already conjectured in Ref.~I that there
should be a localization--delocalization transition with increasing
number of scatterers. In the present work further arguments are given
that for a sufficiently large number of scatterers there is indeed a
mobility edge, separating the band from the tail states, where such a
transition takes place. These effect is accompanied by a considerable
depletion of the density of states. There is a fundamental difference
to the $\sigma$--model which will be discussed.

The article is organized as follows. In Section~\ref{sec2} the main
results of Ref.~I are recapitulated and the $k$--point correlation
function is calculated generalizing a method developed in reference
\cite{Gu1}. In Section~\ref{sec3} the results are specialized to the
strong coupling limit, and it is shown that everywhere within the band
random--matrix results are recovered. In Section~\ref{sec4} the
behaviour of the $k$--point correlation close to the band edge is
studied. The two--point correlation function in particular shows a
transition from GUE behaviour to that of completely uncorrelated
eigenvalues suggesting that there is indeed a mobility edge.

\section{The Model and its Supersymmetric Evaluation}
\label{sec2}

We setup the model in Sec.~\ref{sec2.1} and map it onto superspace in
Sec.~\ref{sec2.2}. The kernel determining all correlation functions is
calculated exactly in Sec.~\ref{sec2.3}. The density of states is
worked out in Sec.~\ref{sec2.4} in the strong coupling limit.

\subsection{Setup of the Model}
\label{sec2.1}

In Ref.~I the density of states was calculated for a billiard with
randomly distributed scatterers, averaged of the the positions of the
scatterer. The system was described by the Hamiltonian
\begin{equation}\label{eq1.01}
  H=H_0+V\,,
\end{equation}
where $H_0$ is the operator of kinetic energy, and $V$ is the
scattering potential.  Assuming $L$ point--like scatterers at
positions $\vec{r}_l$, we have
\begin{equation}\label{eq1.02}
V(\vec{r})=4\pi\lambda\sum\limits_{l=0}^L\delta(\vec{r}-\vec{r}_l) \ .
\end{equation}
Using standard supersymmetric techniques, the density of states was
expressed as the derivative
\begin{equation}\label{eq1.03}
  \rho(E)=\frac{1}{2\pi}\left.\frac{\rmd}{\rmd J}
  \im\left\langle Z(E+J,E-J)\right\rangle\right|_{J=0}\,,
\end{equation}
of the generating function
\begin{eqnarray}
\label{eq1.04}
\fl \lefteqn{Z(E_1,E_2)=}\\\nonumber
\fl\qquad\int \rmd[x]\, \exp\left(\rmi
  \sum\limits_{\alpha\beta}
  \left[\left(E_{1+}\delta_{\alpha\beta}-(H_0)_{\alpha\beta}\right)x_\alpha^*x_\beta+
  \left(E_{2+}\delta_{\alpha\beta}-(H_0)_{\alpha\beta}\right)\xi_\alpha^*\xi_\beta\right]\right)
  M^L\,,
\end{eqnarray}
with the volume element $\rmd[x]=\prod\limits_{\alpha=1}^N \rmd
x_{\alpha}^*\rmd x_{\alpha}\rmd\xi_{\alpha}^* \rmd\xi_{\alpha}$.
Here, the quantity $M$ is given by
\begin{equation}\label{eq1.05}
  M=\left<\exp\left(-4\pi\rmi\lambda\sum\limits_{\alpha\beta}
  \psi_\alpha^*(r)\psi_\beta(r)
  \left(x_\alpha^* x_\beta+\xi_\alpha^* \xi_\beta\right)
  \right)\right> \ ,
\end{equation}
where the brackets denote the average over the scatterer position. To
perform the average, in Ref.~I a trick was applied by replacing the
average over the positions by an integral over the wave function
amplitudes $\psi$ at the positions of the scatterers with the
amplitude probability density $p(\psi)$ as a weight function. For the
latter a Gaussian distribution was taken typically for chaotic
billiards \cite{ber77a,McD}. In the next step, a second average was
performed by replacing the billiard spectrum with that of a random
matrix from the GUE of rank $N$.

As a result a simple analytic expression was obtained for the density
of states. For $L>N$ a qualitative change in the density of states was
observed suggesting a localization--delocalization transition. In the
following the results of Ref.~I will be generalized to the calculation
of the $k$--point correlation function, and further evidence will be
presented of the existence of localized states and a certain type of
mobility edge within the present model.

\subsection{Supersymmetric Matrix Model}
\label{sec2.2}

To compute the $k$--level correlation functions of $k$ energies $E_p,
\ p=1,\ldots,k$, we combine and extend the procedures outlined in
Refs.~I and~\cite{Gu1}.  We construct the functions
$\widehat{R}_k(E_1,\ldots,E_k)$ obtained from averaging over the
product of $k$ Green functions, including their real parts. For
example, the density defined in Eq.~(\ref{eq1.03}) is the imaginary
part of $\widehat{R}_1(E)$. The correlation functions
$R_k(E_1,\ldots,E_k)$ for the imaginary parts only can be
calculated as proper linear combinations from the functions
$\widehat{R}_k(E_1,\ldots,E_k)$. The latter are given as the
derivatives
\begin{eqnarray}
\widehat{R}_k(E_1,\ldots,E_k) = \frac{1}{(2\pi)^k}
                                \prod_{p=1}^k\frac{\partial}{ \partial J_p}
                                \langle Z_k(E+J)\rangle \Bigg|_{J=0}
\label{eq2.31}
\end{eqnarray}
of the generating function
\begin{eqnarray}
\langle Z_k(E+J)\rangle &=& 2^{k(k-1)} \int d[S]
                            \exp\left(\frac{N}{2\pi^2}\Tr S^2
                                    +\Tr S(E+J)\right)
                                    \nonumber\\
       & & \qquad\qquad\qquad \frac{\Det^NS}{\Det^L(1_{2k}+\lambda S)}
\label{eq2.32}
\end{eqnarray}
with respect to $k$ source variables $J_p, \ p=1,\ldots,k$. Energies
and sources variables are ordered in diagonal matrices
$E=\diag(E_1,E_1,\ldots,E_k,E_k)$ and
$J=\diag(-J_1,+J_1,\ldots,-J_k,+J_k)$.  The generating
function~(\ref{eq2.32}) is the straightforward extension of the
generating function used in Ref.~I to arbitrary $k$. To keep
with the notation in Ref.~I, we introduced the $2k\times 2k$
Hermitean supermatrix $S$ which can be mapped onto the supermatrix
$\sigma$ used in Ref.~\cite{Gu1} by exchanging its bosonic and
fermionic eigenvalues. Moreover, we use the symbols $\Tr$ and $\Det$
to indicate supertrace and superdeterminant.

As in Ref.~\cite{Gu1}, the supersymmetric extension of the
Itzykson--Zuber integral can be employed to reduce the generating
function to an integral over the fermionic and bosonic eigenvalues
$is_{p2}, \ p=1,\ldots,k$ and $s_{p1}, \ p=1,\ldots,k$, respectively.
This is so because the term coupling $S$ and $E+J$ is the only one in
the integrand which is not invariant under an unitary transformation
of $S$. Again, as in Ref.~\cite{Gu1}, an easy evaluation of the
derivatives with respect to the source variables is possible and we
arrive at
\begin{eqnarray}
\widehat{R}_k(E_1,\ldots,E_k) &=& \frac{1}{(-\pi^2)^k}
    \int d[s] B_k(s) \exp\left(\frac{N}{2\pi^2}\Tr s^2 +\Tr sE\right)
                                    \nonumber\\
    & & \qquad\qquad \frac{\Det^Ns^+}{\Det^L(1_{2k}+\lambda s)} \, .
\label{eq2.33}
\end{eqnarray}
We collect the eigenvalues in the diagonal matrix
$s=\diag(is_{12},\ldots,is_{k2},s_{11},\ldots,s_{k1})$. We notice that
the bosonic eigenvalues carry a small imaginary increment,
$s_{p1}^+=s_{p1}+i\eta$, where it is necessary. It is send to zero at
an appropriate point of the calculation. Keeping this in mind, we can
integrate all eigenvalues over the entire real axis. This is
equivalent to the choice of the integration contour in Ref.~I. The
function $B_k(s)$
\begin{eqnarray}
B_k(s) = \det\left[\frac{1}{s_{p1}-is_{q2}}\right]_{p,q=1,\ldots,k}
\label{eq2.34}
\end{eqnarray}
in Eq.~(\ref{eq2.32}) is the square root of the Jacobian which is due
to the change of variables from the Cartesean coordinates in $S$ to
eigenvalues $s$ and angles. It is a determinant which couples one
bosonic and one fermionic eigenvalue in each of its
elements. Expanding the determinant,
\begin{equation}\label{eq2.32a}
  B_k(s)=\sum\limits_\pi \varepsilon(\pi)\prod\limits_{l=1}^k
  \frac{1}{s_{pl}-is_{q\pi(l)}}\,,
\end{equation}
where the sum is over all permutations $\pi$, and
$\varepsilon(\pi)=\pm 1$ for even, and odd permutations, respectively,
the integrations in equation (\ref{eq2.33}) factorize into products of
double integrals, each over one bosonic and one fermionic
variable. The result can again be written in terms of a determinant

\begin{eqnarray}
\widehat{R}_k(E_1,\ldots,E_k) =
       \det\left[\widehat{C}_{NL}(E_p,E_q)\right]_{p,q=1,\ldots,k} \,,
\label{eq2.35}
\end{eqnarray}
with a kernel given by
\begin{eqnarray}
\widehat{C}_{NL}(E_p,E_q) &=& -\frac{1}{\pi^2}
                     \int_{-\infty}^{+\infty}
                     \int_{-\infty}^{+\infty} \frac{ds_1ds_2}{s_1-is_2}
                     \exp\Biggl(-\frac{N}{2\pi^2}(s_1^2+s_2^2)\Biggr.
                                    \nonumber\\
              & & \qquad\qquad
                                    \Biggl. +is_2E_q-s_1E_p\Biggr)
                 \left(\frac{1+\lambda s_1}{1+\lambda is_2}\right)^L
                 \left(\frac{is_2}{s_1^+}\right)^N \, .
\label{eq2.36}
\end{eqnarray}
We suppress the indices $p$ and $q$ in the integration
variables. Thus, the correlation functions have a determinant
structure which is a immediate consequence of the
determinant~(\ref{eq2.34}). In full analogy to Ref.~\cite{Gu1}, we
obtain the correlation functions $R_k(E_1,\ldots,E_k)$ by replacing
$1/(s_1^+)^N$ in Eq.~(\ref{eq2.36}) with its imaginary part $\im
1/(s_1^+)^N$.  As in Ref.~I, we rescale the energies and the strength
parameter according to
\begin{eqnarray}
\varepsilon_p = \frac{\pi}{\sqrt{2N}} E_p \qquad {\rm and} \qquad \alpha =
\frac{\sqrt{N/2}}{\pi\lambda} \,. \label{eq2.39}
\end{eqnarray}
On this scale, the correlation functions are given by
\begin{eqnarray}
R_k(\varepsilon_1,\ldots,\varepsilon_k) =
\det\left[C_{NL}(\varepsilon_p,\varepsilon_q)\right]_{p,q=1,\ldots,k} \ ,
\label{eq2.41}
\end{eqnarray}
where the kernel now reads
\begin{eqnarray}
C_{NL}(\varepsilon_p,\varepsilon_q) &=& -\frac{1}{\pi^2}
                     \int_{-\infty}^{+\infty}
                     \int_{-\infty}^{+\infty} \frac{ds_1ds_2}{s_1-is_2}
                                    \nonumber\\
              & & \qquad\qquad
                     \exp\left(-(s_1^2+s_2^2)
                     +2is_2\varepsilon_q-2s_1\varepsilon_p\right)
                                    \nonumber\\
              & & \qquad\qquad
                 \left(\frac{\alpha + s_1}{\alpha + is_2}\right)^L
                 (is_2)^N \im \frac{1}{(s_1^+)^N} \,.
\label{eq2.42}
\end{eqnarray}
Due to the rescaling~(\ref{eq2.39}), we obtain the kernel for the GUE
correlation functions exactly in the form given in Ref.~\cite{Gu1}, if
we consider the limit $\lambda\to 0$, i.e.~$\alpha\to\infty$, or,
equivalently, $L=0$. We notice that the scaling factor in
Eq.~(\ref{eq2.39}) is precisely the GUE mean level spacing
$\pi/\sqrt{2N}$ in the center of the semicircle.

In this derivation, we have omitted a Efetov--Wegner or Rothstein
contribution~\cite{efe83,VWZ,Roth,Gu1} which adds to the real part of
$\widehat{C}_{NL}(E_p,E_q)$. The functions $C_{NL}(E_p,E_q)$, the main
objects of our interest, are not affected.

\subsection{Exact Computation of the Kernel}
\label{sec2.3}

Extending the methods of Ref.~\cite{Gu1}, the kernel can be
evaluated exactly for all values of $N$, $L$ and $\alpha$.
We define the functions
\begin{eqnarray}
u_{NL}(\varepsilon) &=& \frac{(-2i)^N\alpha^L}{\sqrt{\pi}}
                        \int_{-\infty}^{+\infty} ds_2
                        \exp\left(-(s_2-i\varepsilon)^2\right)
                        \frac{s_2^N}{(\alpha+is_2)^L}
                                    \nonumber\\
v_{NL}(\varepsilon) &=& \frac{(-1)^{N+1}N!}{\pi\alpha^L}
                        \exp\left(\varepsilon^2\right)
                                    \nonumber\\
                    & & \qquad
                        \int_{-\infty}^{+\infty} ds_1
                        \exp\left(-(s_1+\varepsilon)^2\right)
                        (\alpha+s_1)^L
                        \im \frac{1}{(s_1^+)^{N+1}}
                                    \nonumber\\
                    &=& \frac{(-1)^N}{\alpha^L}
                        \exp\left(\varepsilon^2\right)
                        \frac{\partial^N}{\partial s_1^N}
                        \exp\left(-(s_1+\varepsilon)^2\right)
                        (\alpha+s_1)^L\Bigg|_{s_1=0} \ ,
\label{eq2.44}
\end{eqnarray}
which reduce to the Hermite polynomials $H_N(\varepsilon)$ for $L=0$
or, equivalently, for $\alpha\to\infty$. In~\ref{app1}, some
properties of these functions are compiled.  We now express $\im
1/(s_1^+)^N$ in Eq.~(\ref{eq2.42}) as
$\partial^{N-1}\delta(s_1)/\partial s_1^{N-1}$ and integrate by parts
until the $(N-1)$--fold derivative with respect to $s_1$ acts on all
$s_1$ dependent terms in the integrand. After applying Leibniz' rule
for multiple derivatives of products, we can insert the second form of
the function $v_{NL}(\varepsilon)$ into Eq.~(\ref{eq2.42}). The $s_2$
integration then yields just the function $u_{NL}(\varepsilon)$ and we
arrive at
\begin{eqnarray}
C_{NL}(\varepsilon_p,\varepsilon_q) = \frac{1}{\sqrt{\pi}}
         \exp\left(-\varepsilon_q^2\right)
         \sum_{n=0}^{N-1} \frac{1}{2^nn!}
         v_{nL}(\varepsilon_p) u_{nL}(\varepsilon_q) \,.
\label{eq2.45}
\end{eqnarray}
Thus, we have expressed the kernel and all correlations in terms of
the functions $v_{NL}(\varepsilon_p)$ and
$u_{NL}(\varepsilon_q)$. Formula~(\ref{eq2.45}) is a generalization of
the corresponding expression for the GUE.  We mention in passing that
one also derives
\begin{eqnarray}
C_{NL}(\varepsilon_p,\varepsilon_q) &=&
         \frac{(-1)^{N-1}}{2^{N-1}(N-1)!\sqrt{\pi}}
                                   \nonumber\\
       & & \int_0^\infty \exp\left(-(\varepsilon_q+t)^2\right)
         u_{NL}(\varepsilon_q+t) v_{(N-1)L}(\varepsilon_p+t) dt \,,
\label{eq2.46}
\end{eqnarray}
which again generalizes the corresponding expression for the GUE in
Ref.~\cite{Gu1}. The result~(\ref{eq2.46}) involves only the orders
$N$ and $N-1$ of the functions $v_{NL}(\varepsilon_p)$ and
$u_{NL}(\varepsilon_q)$, which are not even orthogonal.

\subsection{A Christoffel--Darboux Formula for
            $C_{NL}(\varepsilon_p,\varepsilon_q)$}
\label{sec2.4}

For $L=0$ or, alternatively, $\alpha\to\infty$, the sum on the right
hand side of expression (\ref{eq2.45}) can be performed with the
result
\begin{eqnarray}
C_{N0}(\varepsilon_p,\varepsilon_q)&=&
  \frac{1}{2^{N-1}(N-1)!\sqrt{\pi}}
\exp\left(-\varepsilon_q^2\right)\nonumber\\
        & &\times
        \frac{u_{(N-1)0}(\varepsilon_q)v_{N0}(\varepsilon_p)-u_{N0}
         (\varepsilon_q)v_{(N-1)0}(\varepsilon_p)}
             {\varepsilon_p-\varepsilon_q} \,.
\label{eq2a.01}
\end{eqnarray}
This is the well--known Christoffel--Darboux formula for the Hermite
polynomials. This expression is now generalized to arbitrary values of
$L$. To this end we multiply both sides of equation (\ref{eq2.42}) by
$\varepsilon_p-\varepsilon_q$ and obtain in a sequence of elementary
steps, including one integration by parts,
\begin{eqnarray}
\fl (\varepsilon_p-\varepsilon_q)
     C_{NL}(\varepsilon_p,\varepsilon_q) &=&
     -\frac{1}{\pi^2}\im
                     \int_{-\infty}^{+\infty}
                     \int_{-\infty}^{+\infty} \frac{ds_1ds_2}{s_1-is_2}
                     (\varepsilon_p-\varepsilon_q)
                                    \nonumber\\
     & &\times       \exp\left(-(s_1^2+s_2^2)
                     +2is_2\varepsilon_q-2s_1\varepsilon_p\right)
                 \left(\frac{\alpha + s_1}{\alpha + is_2}\right)^L
                 \left(\frac{is_2}{s_1^+}\right)^N \nonumber\\[1ex]
              &=& -\frac{1}{\pi^2}\im
                     \int_{-\infty}^{+\infty}
                     \int_{-\infty}^{+\infty} \frac{ds_1ds_2}{s_1-is_2}
                            \nonumber\\
     & &\times
   \left[-\frac{1}{2}\left(\frac{\partial}{\partial s_1}+
                \frac{1}{i}\frac{\partial}{\partial s_2}\right)
      \exp\left(2is_2\varepsilon_q-2s_1\varepsilon_p\right)\right]
                                    \nonumber\\
     & &\times \exp\left(-(s_1^2+s_2^2)\right)
         \left(\frac{\alpha + s_1}{\alpha + is_2}\right)^L
                 \left(\frac{is_2}{s_1^+}\right)^N\nonumber\\[1ex]
              &=& -\frac{1}{\pi^2}\im
                     \int_{-\infty}^{+\infty}
                     \int_{-\infty}^{+\infty} \frac{ds_1ds_2}{s_1-is_2}
     \exp\left(2is_2\varepsilon_q-2s_1\varepsilon_p\right)
                                    \nonumber\\
              & &\times
              \left[\frac{1}{2}\left(\frac{\partial}{\partial s_1}+
                \frac{1}{i}\frac{\partial}{\partial s_2}\right)
     \exp\left(-(s_1^2+s_2^2)\right)
     \left(\frac{\alpha + s_1}{\alpha + is_2}\right)^L
       \left(\frac{is_2}{s_1^+}\right)^N\right]\nonumber\\[1ex]
              &=& -\frac{1}{\pi^2}\im
                     \int_{-\infty}^{+\infty}
                     \int_{-\infty}^{+\infty}
                     \frac{ds_1ds_2}{s_1-is_2}\nonumber\\
              & & \times
     \left[\frac{1}{2}\left(-2s_1+2is_2+\frac{L}{\alpha+s_1}
   -\frac{L}{\alpha+is_2}+\frac{N}{is_2}-\frac{N}{s_1^+}\right)\right]
                                    \nonumber\\
              & &\times
              \exp\left(-(s_1^2+s_2^2)
                    +2is_2\varepsilon_q-2s_1\varepsilon_p\right)
              \left(\frac{\alpha + s_1}{\alpha + is_2}\right)^L
                 \left(\frac{is_2}{s_1^+}\right)^N\nonumber\\[1ex]
              &=& -\frac{1}{\pi^2}\im
                     \int_{-\infty}^{+\infty}
                     \int_{-\infty}^{+\infty} ds_1ds_2\nonumber\\
              & & \times
              \frac{1}{2}\left(-2-\frac{L}{(\alpha+s_1)(\alpha+is_2)}
              +\frac{N}{is_2s_1^+}\right)\nonumber\\
              & &\times
              \exp\left(-(s_1^2+s_2^2)
                    +2is_2\varepsilon_q-2s_1\varepsilon_p\right)
              \left(\frac{\alpha + s_1}{\alpha + is_2}\right)^L
                 \left(\frac{is_2}{s_1^+}\right)^N\,.
\label{eq2a.02}
\end{eqnarray}
The inconvenient denominator coupling the $s_1$ and the $s_2$
integrations has disappeared with the consequence that all integrals
can be expressed in terms of the $u_{NL}(\varepsilon)$ and
$v_{NL}(\varepsilon)$. A formula of the Christoffel--Darboux type
obtains
\begin{eqnarray}
(\varepsilon_p-\varepsilon_q)C_{NL}(\varepsilon_p,\varepsilon_q)&=&
\frac{1}{2^N(N-1)!\sqrt{\pi}}\exp\left(-\varepsilon_q^2\right)
           \nonumber\\
        &&\times
        \Big[-u_{NL}(\varepsilon_q)v_{(N-1)L}(\varepsilon_p)
        +u_{(N-1)L}(\varepsilon_q)v_{NL}(\varepsilon_p)\nonumber\\
        &&-\frac{L}{2\alpha^2}
        u_{N(L+1)}(\varepsilon_q)v_{(N-1)(L-1)}(\varepsilon_p)\Big]
          \ ,
\label{eq2a.03}
\end{eqnarray}
which is valid for all values of $N$, $L$ and $\alpha$. This is quite
remarkable, because the functions $u_{NL}(\varepsilon)$ and
$v_{NL}(\varepsilon)$ are no orthogonal polynomials. In different
context, similar generalizations of the Christoffel--Darboux formula
have been obtained in Refs.~\cite{Gu2,Gu3} and by Strahov and
Fyodorov~\cite{SF} in the calculation of correlation functions of
ratios and products of characteristic polynomials of Hermitian random
matrices.

For $L=0$ expression (\ref{eq2a.01}) for the Hermite polynomials is
recovered. Another special case is obtained for the strong coupling
limit. Here the Gauss functions in the integral (\ref{eq2.42}) may be
replaced by one with the consequence that the first term in the
brackets on the right hand side of equation (\ref{eq2a.03}) is
missing.  Furthermore in this limit the $u_{NL}(\varepsilon)$ and
$v_{NL}(\varepsilon)$ can be expressed in terms of generalized
Laguerre polynomials,
\begin{eqnarray}
 u_{NL}(\varepsilon) &=&
   \sqrt{\pi}\exp(\varepsilon^2)(-1)^N(2\alpha)^{(N+1)}
      \nonumber\\
  & & \times
         \frac{N!}{(L-1)!}\exp(-z)z^{L-N-1}L_N^{(L-N-1)}(z)\\
   v_{NL}(\varepsilon) &=&(-1)^NN!\alpha^{-N}L_N^{(L-N)}(z)\ ,
\label{eq2a.04}
\end{eqnarray}
where $z=2\varepsilon\alpha=E/\lambda$ as in Ref.~I. Collecting the
results we obtain from equation~(\ref{eq2a.03})
\begin{eqnarray}
\fl (\varepsilon_p-\varepsilon_q)C_{NL}(\varepsilon_p,\varepsilon_q)&=&
\frac{N!}{(L-1)}\exp\left(-z_q\right)z_q^{L-N}\nonumber\\
        &&\times
        \left[-L_{N-1}^{(L-N)}(z_q)L_N^{(L-N)}(z_p)
        +L_N^{(L-N)}(z_q)L_{N-1}^{(L-N)}(z_p)\right] \ .
\label{eq2a.05}
\end{eqnarray}
Comparing equations (\ref{eq2.45}) and (\ref{eq2a.03}) we obtain the
following Christoffel--Darboux relation for the generalized Laguerre
polynomials
\begin{eqnarray}
\sum_{n=0}^{N-1}n!x^{N-n-1}L_n^{(L-n-1)}(x)L_n^{(L-n)}(y)=\nonumber\\\qquad\qquad
          N!\frac{L_N^{(L-N)}(x)L_{N-1}^{(L-N)}(y)
        -L_{N-1}^{(L-N)}(x)L_N^{(L-N)}(y)}{y-x}          \,.
\label{eq2a.06}
\end{eqnarray}
This is not the Christoffel--Darboux relation for the Laguerre
polynomials found in compilation such as Ref.~\cite{Abram}, but we
cannot exclude that it is known in the mathematical literature.

\section{Density of States and Correlations via a Saddlepoint
         Approximation}
\label{sec3}

In Section~\ref{sec3.1}, we work out the density of states in the
strong coupling limit. The correlations in the bulk of the spectrum
are computed for arbitrary coupling in Section~\ref{sec3.2}.

\subsection{Density of States in the Strong Coupling Limit}
\label{sec3.1}

\begin{figure}
  \centering \includegraphics[width=7.5cm]{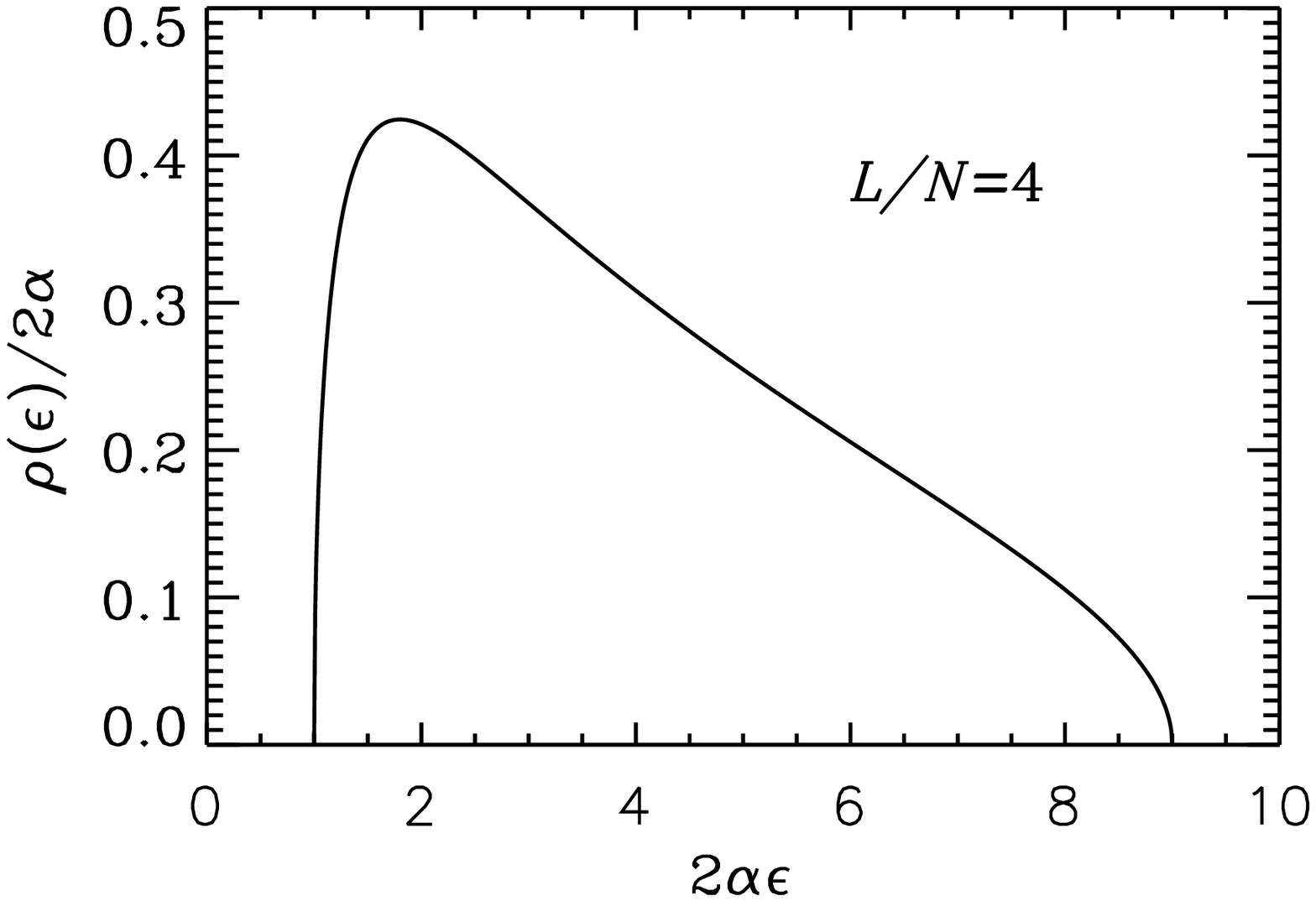}
  \includegraphics[width=7.5cm]{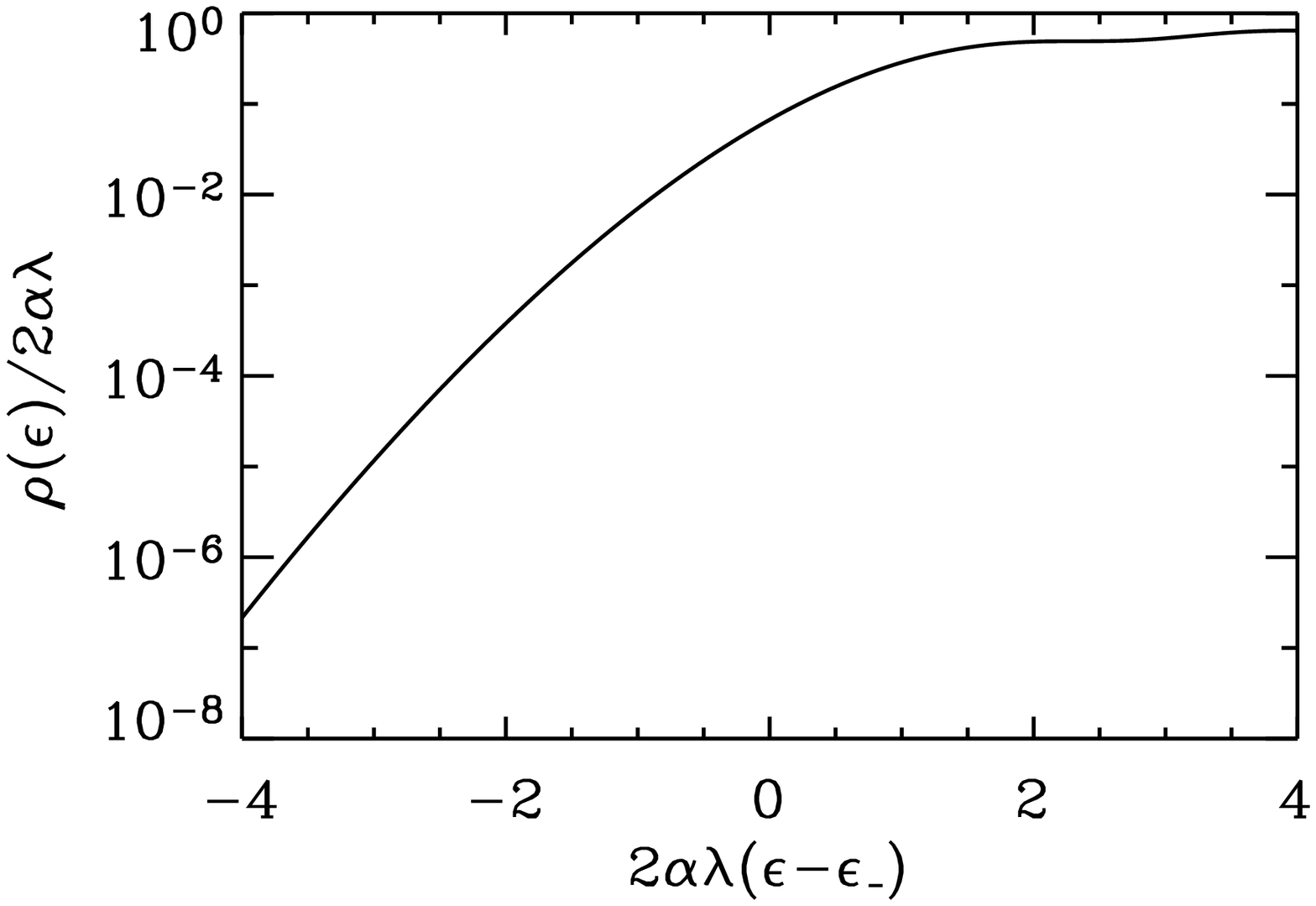}
  \caption{(a) Density of
  states in the strong coupling limit for $L/N=4$.  (b)~Density of
  states in the neighbourhood of the lower band edge (for details see
  section \ref{sec4.2}).}\label{fig1}
\end{figure}

For strong coupling $\lambda\gg 1$ or $\alpha\ll 1$ and $L>N$, the
density of states was evaluated in Ref.~I by means of a WKB
approximation to leading order in $L>N\gg 1$. Here we show that this
is equivalent to a saddlepoint approximation. For $k=1$, we write the
generating function~(\ref{eq2.45}) in the form
\begin{eqnarray}
\langle Z_1(\varepsilon+J)\rangle &=& \int d[S]
                \exp\left({\cal L}(S,\varepsilon+J)\right)
                                    \nonumber\\
{\cal L}(S,\varepsilon+J) &=& \Tr S^2 + 2\Tr S(\varepsilon+J)
                                 +N\Tr\ln S - L\Tr\ln(\alpha+S) \,.
\label{eq2.51}
\end{eqnarray}
We use the rescaled variables~(\ref{eq2.39}), drop the index 1 on the
energy variable and write $\varepsilon$ shorthand for $\varepsilon
1_2$. In the strong coupling limit $S$ is of the order of $\alpha$ as
can be seen by applying the substitution $S=\alpha S'$. The term $\Tr
S^2$ is of thus of the order of $\alpha^2$ and may be dropped
in the Lagrangean ${\cal L}(S,\varepsilon+J)$. In this approximation, the saddlepoint
equation resulting from the condition $d{\cal L}=0$ at $J=0$ reads
\begin{eqnarray}
  2\varepsilon+\frac{N}{s_0}-\frac{L}{\alpha+s_0} = 0 \,,
\label{eq2.52}
\end{eqnarray}
with $s_0$ standing for the two scalar saddlepoints $s_{10}$
and $is_{20}$. The solutions can be written as
\begin{eqnarray}
s_0=\frac{1}{4\varepsilon}\left(-(2\varepsilon\alpha-(L-N))\mp
           i\sqrt{4LN-(2\varepsilon\alpha-(L-N))^2}\right) \,.
\label{eq2.53}
\end{eqnarray}
Obviously, the imaginary part is only non--zero if the energy
satisfies
\begin{eqnarray}
\varepsilon_- \le \varepsilon \le \varepsilon_+
 \qquad {\rm with} \qquad
\varepsilon_\mp = \frac{1}{2\alpha}\left(L+N\mp 2\sqrt{LN}\right) \,.
\label{eq2.54}
\end{eqnarray}
We now expand the Lagrangean ${\cal L}(S,\varepsilon+J)$ around the
saddlepoints up to second order and integrate out the massive modes in
a Gaussian fashion. One can convince oneself in a straightforward, but
tedious, calculation that these Gaussian integrals converge as long as
the condition~(\ref{eq2.54}) holds.  At the saddlepoints, the
Lagragean is simply $4s_0J_1$ and we find from Eq.~(\ref{eq2.31})
\begin{eqnarray}
\widehat{R}_1(\varepsilon) &=& \frac{2}{\pi} s_0
                        \nonumber\\
                           &=&
\frac{1}{2\pi\varepsilon}\left(-(2\varepsilon\alpha-(L-N)) +
           i\sqrt{4LN-(2\varepsilon\alpha-(L-N))^2}\right) \,.
\label{eq2.55}
\end{eqnarray}
This is the full one--point function in the strong coupling limit. The
imaginary part is the density of states which is non--zero for
$\varepsilon_- \le \varepsilon \le\varepsilon_+$. As expected, it
coincides with the WKB approximation of Ref.~I.  The saddlepoint
approximation yields, in addition, also the real part of the
one--point function. As an illustration figure \ref{fig1}(a) shows the
density of states calculated from the imaginary part of
$\widehat{R}_1(\varepsilon)$ for $L/N=4$.

\subsection{Correlations in the Bulk of the Spectrum}
\label{sec3.2}

We take advantage of a remarkable identity which connects the kernel
and the generating function for $k=1$,
\begin{eqnarray}
\widehat{C}_{NL}(\varepsilon_p,\varepsilon_q) &=&
  -\frac{1}{\pi(\varepsilon_p-\varepsilon_q)}
   \langle Z_1({\cal E})\rangle
                     \nonumber\\
   &=& -\frac{1}{\pi(\varepsilon_p-\varepsilon_q)}
       \int d[S] \exp\left(\Tr S^2 +\Tr S{\cal E}\right)
                      \nonumber\\
   & & \qquad\qquad   \frac{\Det^NS}{\Det^L(\alpha1_2+S)}
\label{eq2.61}
\end{eqnarray}
with ${\cal E}=\diag(\varepsilon_p,\varepsilon_q)$. This identity
which is easily derived with the supersymmetric extension of the
Itzykson--Zuber integral for $k=1$ allows us to work out the
correlations on the unfolded scale by a saddlepoint approximation
involving $2\times 2$ supermatrices, i.e.~in a simple Cartesian
space. A similar procedure was employed in Ref.~\cite{GW} in the
context of chiral random matrix ensembles.

To begin with, we discuss the strong coupling limit of the previous
section and turn to the general case later. We write
\begin{eqnarray}
{\cal E} &=& \varepsilon_{pq}1_2 + D\frac{r_{pq}}{2}\Lambda
\qquad {\rm with} \qquad
\nonumber\\
 \varepsilon_{pq} &=& \frac{\varepsilon_p+\varepsilon_q}{2}
 \qquad {\rm and} \qquad Dr_{pq} =
 \varepsilon_p-\varepsilon_q \ .
\label{eq2.62}
\end{eqnarray}
Here, we introduced the metric $\Lambda=\diag(+1,-1)$ and,
anticipating the steps to come, the local mean level spacing
$D=1/R_1(\varepsilon_{pq})$ which defines the unfolded scale. We use
the form~(\ref{eq2.51}) for the generating function with
$\varepsilon+J$ replaced by ${\cal E}$ to evaluate
Eq.~(\ref{eq2.61}). In the strong coupling limit, we neglect the term
$\Tr S^2$. Although $\alpha\ll 1$ in this limit, we do not make any
assumption about its value in the present discussion. The saddlepoints
are the stable points of the integrand in an asymptotic $1/N$
expansion. The unfolded correlations live on the local scale of the
mean level spacing $D$. Thus, we have to keep
$r_{pq}=(\varepsilon_p-\varepsilon_q)/D$ fixed in the asymptotic
expansion for the calculation of the unfolded correlations. The energy
difference $\varepsilon_p-\varepsilon_q$ itself appears in the
integrand. As it is given by $Dr_{pq}$, and as the mean level spacing
$D$ vanishes in the limit $N\to\infty$, the energy difference cannot
yield a contribution to the saddlepoints and we may neglect it when
calculating them. Thus, we are left with exactly the same problem as
in the previous section, only $\varepsilon$ is replaced by
$\varepsilon_{pq}$. This implies that the integrals over the massive
modes converge in the non--zero region of the spectrum and, moreover,
that the only non--vanishing contribution to the correlations comes
from the term $\Tr S{\cal E}$ in the Lagrangean. Collecting
everything, we find
\begin{eqnarray}
\widehat{c}_{NL}(r_{pq},\widetilde{r}_{pq}) =
\lim_{L>N\to\infty} D \widehat{C}_{NL}(\varepsilon_p,\varepsilon_q)
 = \exp(\pi\widetilde{r}_{pq})
     \frac{\exp(i\pi r_{pq})}{\pi r_{pq}}
\label{eq2.63}
\end{eqnarray}
for the kernel on the unfolded scale. We notice that the result
depends on $\widetilde{r}_{pq}=(\varepsilon_p-\varepsilon_q)
\widetilde{R_1}(\varepsilon_{pq})$ where $\widetilde{R_1}(\varepsilon_{pq})=\re\widehat{R_1}(\varepsilon_{pq})$ is the real part of the
one--point function. As discussed in Ref.~\cite{Gu2}, an
Efetov--Wegner or Rothstein term has to be added to
Eq.~(\ref{eq2.63}). It affects only the real part and reads $-1/\pi
r_{pq}$. For the correlation functions involving the imaginary parts
of the Green functions, we only need the imaginary part
\begin{eqnarray}
c_{NL}(r_{pq},\widetilde{r}_{pq}) =
\im \widehat{c}_{NL}(r_{pq},\widetilde{r}_{pq})
   = \exp(\pi\widetilde{r}_{pq})
     \frac{\sin\pi r_{pq}}{\pi r_{pq}} \,,
\label{eq2.64}
\end{eqnarray}
which consists of the GUE sine kernel and an exponential function
depending on $\widetilde{r}_{pq}$. Both variables,
$\widetilde{r}_{pq}$ and $r_{pq}$, are odd under the exchange of the
indices $p$ and $q$. Thus, the sine kernel stays unchanged, while
the exponential function acquires a sign in its argument.
This implies for the correlation function on the local scale
\begin{eqnarray}
X_k(r_{12},r_{13},\ldots,r_{(k-1)k}) &=&
\lim_{L>N\to\infty} D^k R_k(\varepsilon_1,\ldots,\varepsilon_k)
               \nonumber\\
 &=&
\det\left[\frac{\sin\pi r_{pq}}{\pi r_{pq}}\right]_{p,q=1,\ldots,k} \,, \label{eq2.65}
\end{eqnarray}
which is identical to the standard GUE correlations.

The previous derivation is for the strong coupling limit. In the
following, we present a general discussion of the correlations. We
write the kernel as the convolution
\begin{eqnarray}
\widehat{C}_{NL}(\varepsilon_p,\varepsilon_q) &=& \frac{1}{\pi}
  \int_{-\infty}^{+\infty} dy_p \exp\left(-y_p^2\right)
  \int_{-\infty}^{+\infty} dy_q \exp\left(-y_q^2\right)
                                    \nonumber\\
& & \qquad\qquad
       \widehat{B}_{NL}(\varepsilon_p+y_p,\varepsilon_q+iy_q)
                                    \nonumber\\
\widehat{B}_{NL}(z_p,z_q) &=& -\frac{1}{\pi^2}
                     \int_{-\infty}^{+\infty}
                     \int_{-\infty}^{+\infty} \frac{ds_1ds_2}{s_1-is_2}
                     \exp\Biggl(-(s_1^2+s_2^2)+is_2z_q-s_1z_p\Biggr)
                                    \nonumber\\
              & & \qquad\qquad
                 \left(\frac{\alpha+s_1}{\alpha+is_2}\right)^L
                 \left(\frac{is_2}{s_1^+}\right)^N \,.
\label{eq2.66}
\end{eqnarray}
Obviously, the kernel $\widehat{B}_{NL}(z_p,z_q)$ is the kernel of the
strong coupling limit. However, it emerges due to the
convolution. Thus, we do not need to assume that $L>N$. In the
following, we only assume that both numbers, $L$ and $N$ are large.
Moreover, we make no assumption about $\alpha$.  As we have seen in
the previous discussion, this kernel, here denoted
$\widehat{B}_{NL}(z_p,z_q)$, leads to standard GUE correlations on the
unfolded scale.  One might argue that this does not necessarily carry
over to the present case, because the arguments
$z_p=\varepsilon_p+y_p$ and $z_q=\varepsilon_q+iy_q$ contain the
integration variables $y_p$ and $y_q$. However, as we are only
interested in the fluctuations, we only need to consider the
integration variables on this scale. Thus, we may neglect them for the
determination of the saddlepoints. After assembling things properly,
we arrive at
\begin{eqnarray}
\widehat{c}_{NL}(r_{pq},\widetilde{r}_{pq}) &=&
\lim_{N\to\infty} D \widehat{C}_{NL}(\varepsilon_p,\varepsilon_q)
                                \nonumber\\
 &=& \frac{1}{\pi}
  \int_{-\infty}^{+\infty} dy_p \exp\left(-y_p^2\right)
  \int_{-\infty}^{+\infty} dy_q \exp\left(-y_q^2\right)
                                \nonumber\\
 & & \qquad  \exp\left(\pi
            (\widetilde{r}_{pq}+
                (y_p-iy_q)\widetilde{R_1}(\varepsilon_{pq}))
          \right)
                                \nonumber\\
 & & \qquad      \frac{\exp\left(i\pi
            (r_{pq}+(y_p-iy_q)/D)\right)}
          {\pi(r_{pq}+(y_p-iy_q)/D)} \,.
\label{eq2.67}
\end{eqnarray}
As we are only interested in the imaginary part, we may again ignore
the Efetov--Wegner or Rothstein term. The imaginary part can be
obtained from the difference of a retarded and an advanced Green
function. The two Green functions yield the same kernels, apart from a
sign change in the argument of the exponential function in the
numerator, $\exp\left(\pm i\pi (r_{pq}+(y_p-iy_q)/D)\right)$. Hence,
only the difference of these two exponential functions, the sine
function, enters. This is equivalent to taking the imaginary part of
Eq.~(\ref{eq2.67}) while formally ignoring the imaginary unit coming
with the variable $y_q$. Thus, we find
\begin{eqnarray}
c_{NL}(r_{pq},\widetilde{r}_{pq}) &=&
\im \widehat{c}_{NL}(r_{pq},\widetilde{r}_{pq})
                                \nonumber\\
 &=& \frac{1}{\pi}
  \int_{-\infty}^{+\infty} dy_p \exp\left(-y_p^2\right)
  \int_{-\infty}^{+\infty} dy_q \exp\left(-y_q^2\right)
                                \nonumber\\
 & & \qquad  \exp\left(\pi
            (\widetilde{r}_{pq}+
                (y_p-iy_q)\widetilde{R_1}(\varepsilon_{pq}))
          \right)
                                \nonumber\\
 & & \qquad      \frac{\sin\left(\pi
            (r_{pq}+(y_p-iy_q)/D)\right)}
          {\pi(r_{pq}+(y_p-iy_q)/D)}
                                \nonumber\\
 &=&  \exp(\pi\widetilde{r}_{pq})
     \frac{\sin\pi r_{pq}}{\pi r_{pq}} \,,
\label{eq2.68}
\end{eqnarray}
where the integrals over $y_p$ and $y_q$ were done as in
Ref.~\cite{Gu3}. Hence, the correlations are, once more, of the
standard GUE type. Some comments are in order. First, it should be
clear that the mean level spacing $D$ in the calculation above was
formally the one of the strong coupling limit and has thus to be
smoothly adjusted when going into another regime. Therefore, our line
of arguing is correct only if we are always in the bulk of the
spectrum, i.e.~far away from any possible edges or gaps. Second, the
discussion beyond the strong coupling limit could also be done in a
saddlepoint approximation of the full expression~(\ref{eq2.61}). This,
however, leads to a most inconvenient third order saddlepoint
equation.  In the approach chosen here we avoid this an also gain the
insight that the strong coupling limit and the general case are
related via a convolution. Third, we emphasize that the
connection~(\ref{eq2.61}) between the kernel and the generating
function for $k=1$ simplifies the calculations enormously: the
saddlepoints are isolated, no Goldstone modes occur. Furthermore, all
correlations are treated at once.

\section{The band edges}
\label{sec4}

{}From the pioneering work of Mott and Anderson it is known that in
disordered systems there are no sharp band edges for the density of
states. There is a mobility edge instead separating the delocalized
states in the band from the localized ones in the tails. The
mathematical origin of the band edges is due to the fact that in
dependence of some parameter the two solutions of the saddle point
equation (\ref{eq2.52}) change from complex conjugate to real. This
behaviour is generic, though the present model the band edges are only
an artifact of the finite rank of the matrices. It therefore is
worthwhile to study the regime of the band edges somewhat more in
detail.

After obtaining a WKB approximation for the kernel in
Section~\ref{sec4.1}, we work out density of states and correlations
in Section~\ref{sec4.1}.

\subsection{A WKB Approximation for
            $C_{NL}(\varepsilon_p,\varepsilon_q)$}
\label{sec4.1}

To keep the discussion simple, we again concentrate on the strong
coupling limit.  Starting point is the Christoffel--Darboux relation
(\ref{eq2a.05}) for $C_{NL}(\varepsilon_p,\varepsilon_q)$ holding in
this limit. Using standard relations for the generalized Laguerre
polynomials, it can be written in the alternative form
\begin{eqnarray}
\fl (\varepsilon_p-\varepsilon_q)C_{NL}(\varepsilon_p,\varepsilon_q)&=&
2\alpha\frac{N!}{(L-1)}\exp(-z_q)z_q^{L-N}\nonumber\\ & \times &
\left[{L_N^{(L-N-1)}}'(z_q)L_N^{(L-N-1)}(z_p)
-L_N^{(L-N-1)}(z_q){L_N^{(L-N-1)}}'(z_p)\right] \,, \label{eq4.01}
\end{eqnarray}
where $z_{p/q}=2\alpha\varepsilon_{p/q}$, which is somewhat more
suitable for the present purpose. Following Ref.~I, we write
$L_N^{(L-N-1)}(z)$ as
\begin{equation}\label{eq4.02}
  L_N^{(L-N-1)}(z)=
  \sqrt{\frac{(L-1)!}{N!}}\exp(z/2)z^{-\frac{L-N}{2}}f(z)\,,
\end{equation}
where $f(z)$ is a solution of
\begin{eqnarray}\label{eq4.03}
  f''(z)+q^2(z)f(z)=0\ ,\nonumber\\
  q^2(z)=\frac{N+L}{2z}-\frac{1}{4}+\frac{1-(L-N-1)^2}{4z^2}\,.
\end{eqnarray}
$q^2(z)$ may be written as
\begin{equation}\label{eq4.04}
  q^2(z)=-\frac{1}{4z^2}(z-z_-)(z-z_+)\,,
\end{equation}
where
\begin{equation}\label{eq4.05}
  z_\pm=N+L\pm\sqrt{4NL+2(L-N)}\approx N+L\pm2\sqrt{NL}\,.
\end{equation}
We note that this is the same expression, which was obtained above
from the saddlepoint approximation for the band edges in the strong
coupling limit (see section \ref{sec3.1}).

For $z_-\ll z\ll z_+$ the WKB solution of equation (\ref{eq4.03}) is
given by
\begin{equation}\label{eq4.06}
  f(z)= \sqrt{\frac{1}{\pi q(z)}}\cos\left[Q(z)-\frac{\pi}{4}\right]\ ,
\end{equation}
where
\begin{equation}\label{eq4.06a}
  Q(z)=\int_{z_-}^zq(t)\,dt\,.
\end{equation}
(To be concise we restrict the discussion to the neighbourhood of the
lower edge, but it is straightforward to transfer all results to the
upper edge as well). Inserting this into equation (\ref{eq4.01}), we
recover the result for $C_{NL}(\varepsilon_p,
\varepsilon_q)$ obtained in section \ref{sec3.2}
by means of the saddlepoint technique.  For $z\ll z_-$ the
corresponding expression reads
\begin{equation}\label{eq4.07}
 f(z)= \frac{1}{2}\sqrt{\frac{1}{\pi
 |q(z)|}}\exp\left[-\left|Q(z)\right|\right]\,.
\end{equation}
Inserting expression (\ref{eq4.07}) into (\ref{eq4.01}) one notices
that $C_{NL}(\varepsilon_p, \varepsilon_q)$ vanishes within the limits
of the WKB approximation applied. To describe this regime
appropriately, one would have to go to the next WKB order.

We do not proceed further in this direction, but concentrate on the
immediate neighbourhood of the lower edge which is not covered by
equations (\ref{eq4.06}) and (\ref{eq4.07}). Linearizing $q(z)$ close
to $z_-$,
\begin{equation}\label{eq4.08}
  q(z)=\sqrt{\frac{z_+-z_-}{4z_-^2}(z-z_-)}\,,
\end{equation}
equation (\ref{eq4.03}) can be solved with the result
\begin{eqnarray}\label{eq4.09}
  f(z)= \frac{1}{\sqrt{\lambda}}\mathrm{Ai}\left[\lambda(z_--z)\right]\ ,\nonumber\\
  \lambda=\left(\frac{\sqrt{z_+-z_-}}{2z_-}\right)^\frac{2}{3}
  =\left(\frac{(NL)^{1/4}}{N+L-2\sqrt{NL}}\right)^\frac{2}{3}\,,
\end{eqnarray}
where $\mathrm{Ai}(z)$ is the Airy function. With the factor
$\lambda^{-1/2}$ the asymptotic behaviour of the Laguerre polynomials
is reproduced correctly by equation (\ref{eq4.02}). This can be shown
by techniques described e.\,g. in chapter 9.3 of
Ref.~\cite{Mor}. Collecting the results, we obtain from equation
(\ref{eq4.01})
\begin{eqnarray}
\fl C_{NL}(\varepsilon_p,\varepsilon_q)&=&2\alpha
\exp\left(-\frac{z_q-z_p}{2}\right)\left(\frac{z_q}{z_p}\right)^\frac{L-N}{2}\nonumber\\
       && \times \left[
        \frac{f'(z_q)f(z_p)-f(z_q)f'(z_p)}{z_p-z_q}-(L-N)\frac{f(z_p)f(z_q)}{2z_pz_q}\right]
          \,.
\label{eq4.10}
\end{eqnarray}
The second term on the right hand side vanishes for $L\to\infty$,
since $z_p$, $z_q$ are of order $\mathrm{O}(N+L)$, and will be
discarded in the following. Essentially the same approach to describe
the behaviour of correlation functions close to the band edges was
applied by Akemann and Fyodorov~\cite{AF} in the study of
characteristic polynomials.

\subsection{The Density of States
            and the $k$--point Correlation Function}
\label{sec4.2}

The density of states is obtained from equation (\ref{eq4.10})
\begin{eqnarray}\label{eq4.11}
  \rho(\varepsilon)=C_{NL}(\varepsilon,\varepsilon)&=&
  2\alpha\left\{\left[f'(2\alpha\varepsilon)\right]^2
  -f(2\alpha\varepsilon)f''(2\alpha\varepsilon)\right\}\,.
\end{eqnarray}
For the regime close to the lower band edge we obtain by inserting
expression (\ref{eq4.09}) for $f(z)$,
\begin{equation}\label{eq4.14}
  \rho(\varepsilon)=
  2\alpha\lambda\left\{\left[\mathrm{Ai}'(-s)\right]^2
  +s\left[\mathrm{Ai}(-s)\right]^2\right\}\ ,\qquad
  s=2\alpha\lambda(\varepsilon-\varepsilon_-)\,.
\end{equation}
Figure \ref{fig1}(b) shows a plot of the density of states in the
transition regime as obtained from equation (\ref{eq4.14}).

\begin{figure}
  \centering \includegraphics[width=7.5cm]{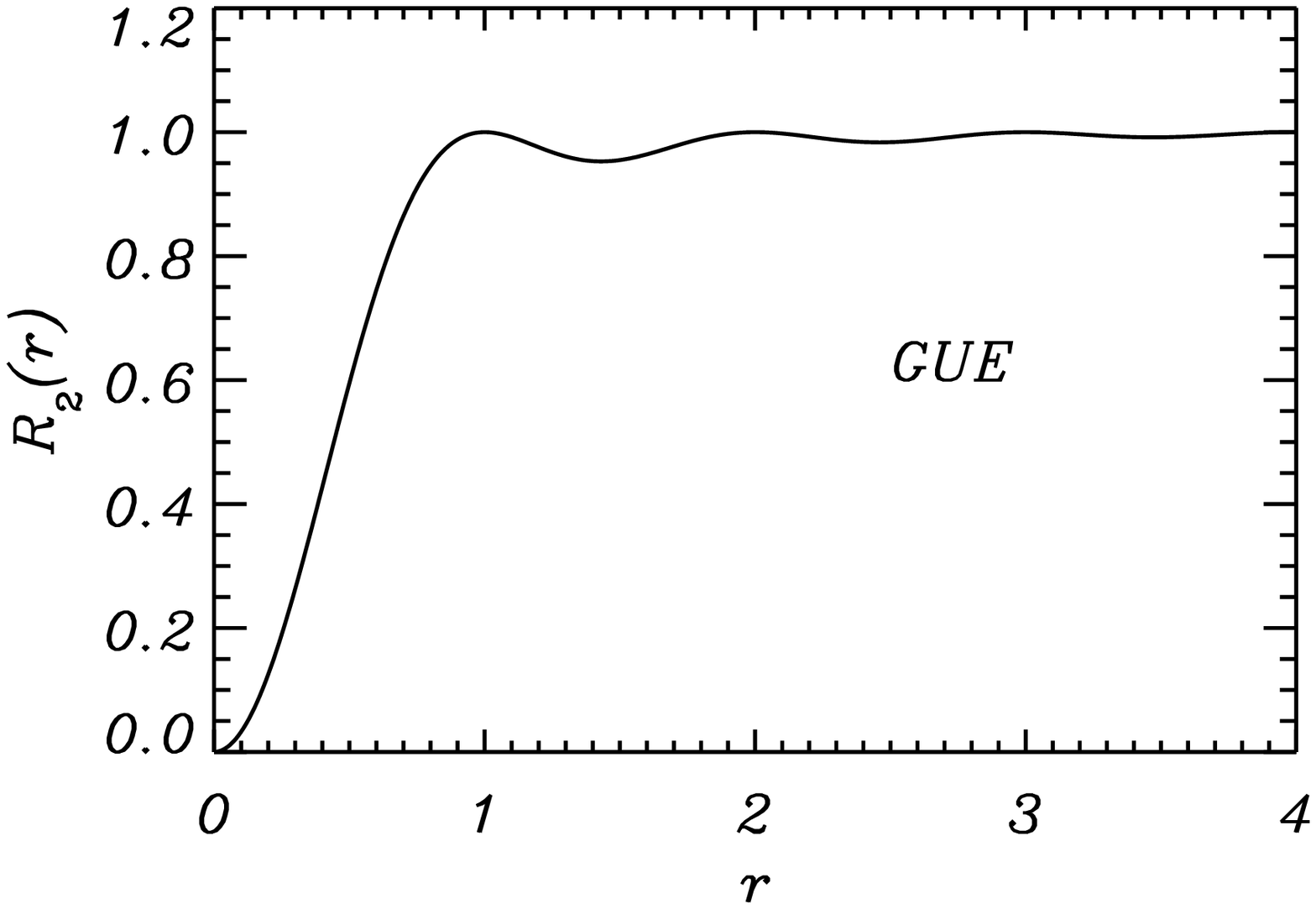}
  \includegraphics[width=7.5cm]{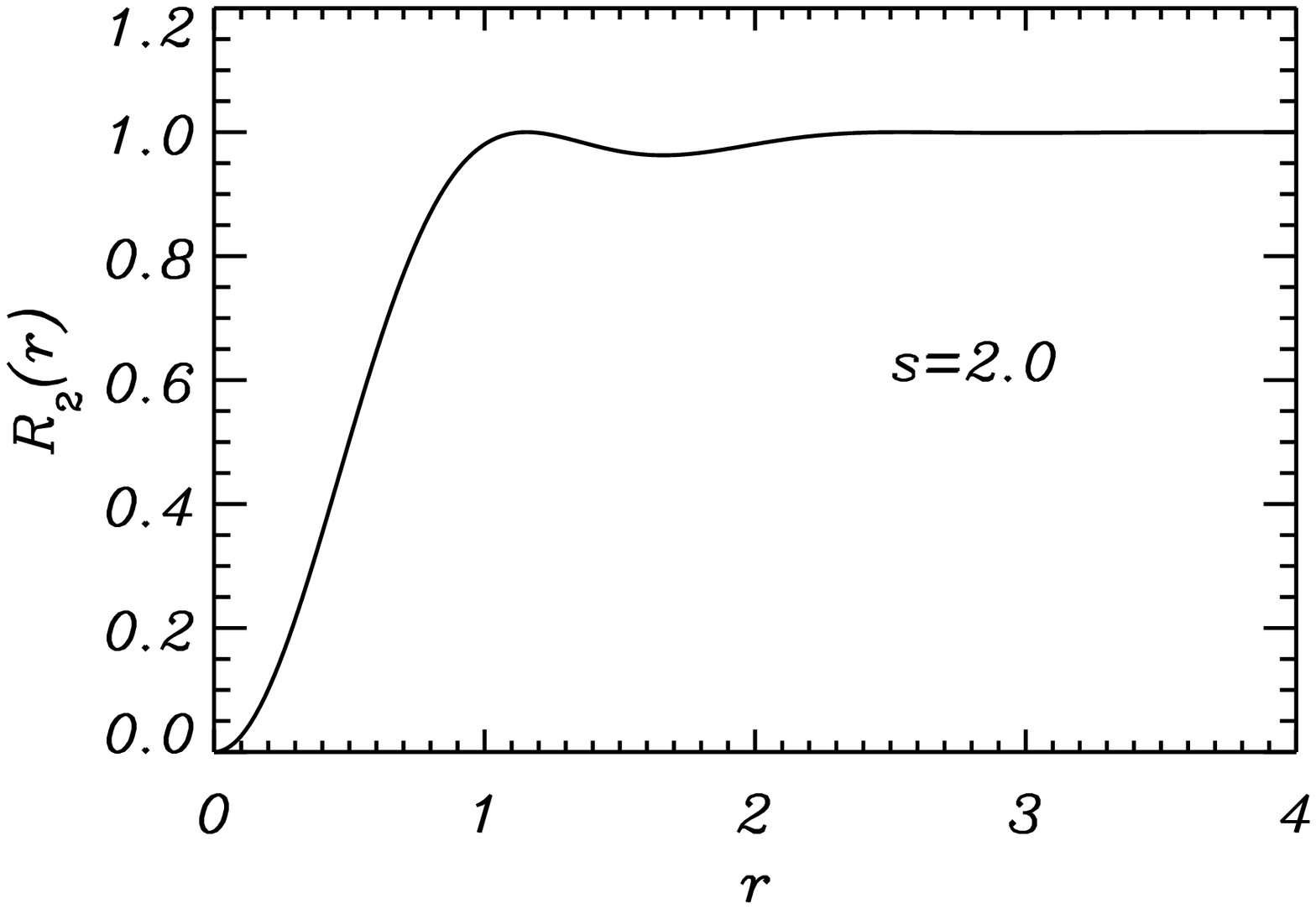}
  \includegraphics[width=7.5cm]{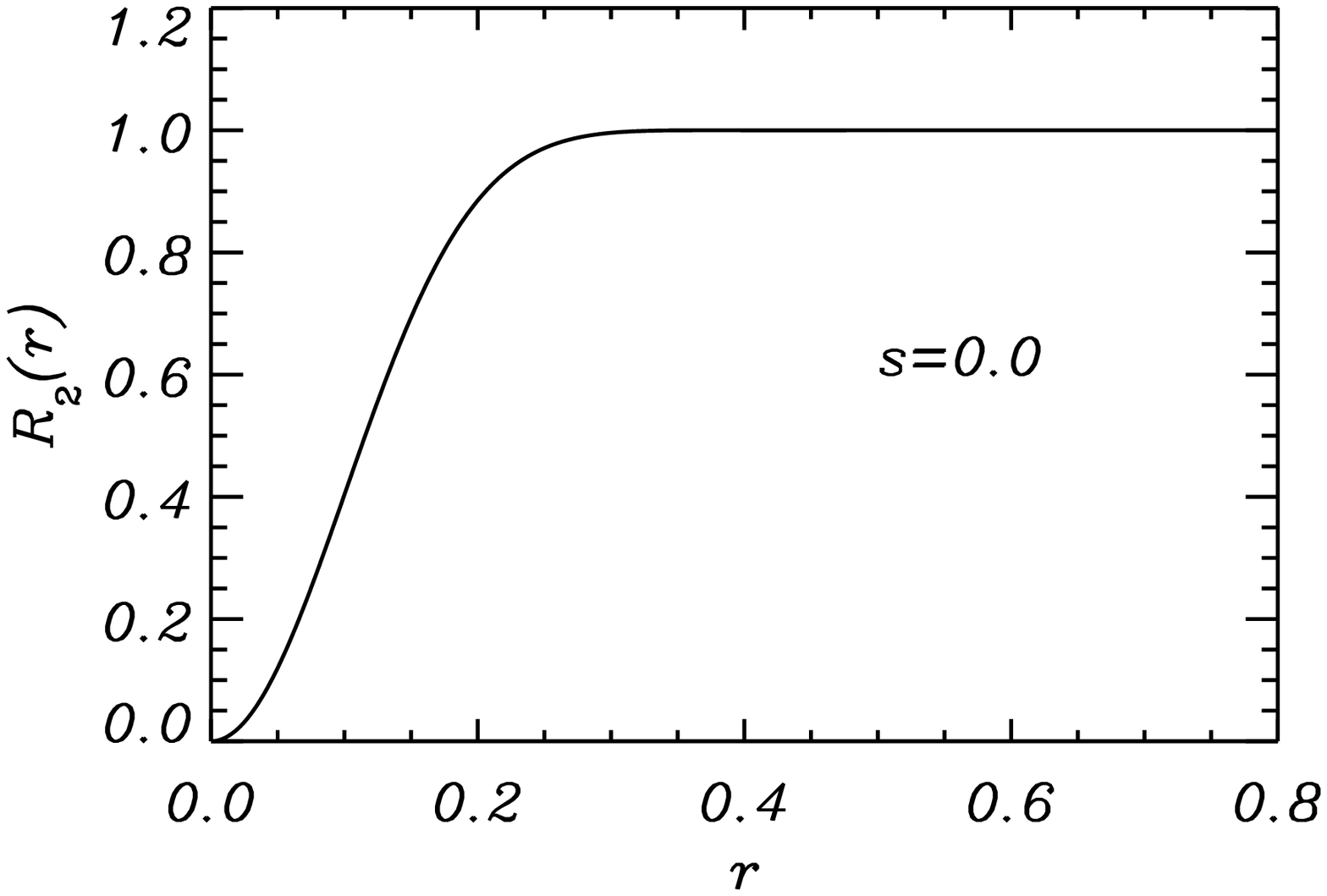}
  \includegraphics[width=7.5cm]{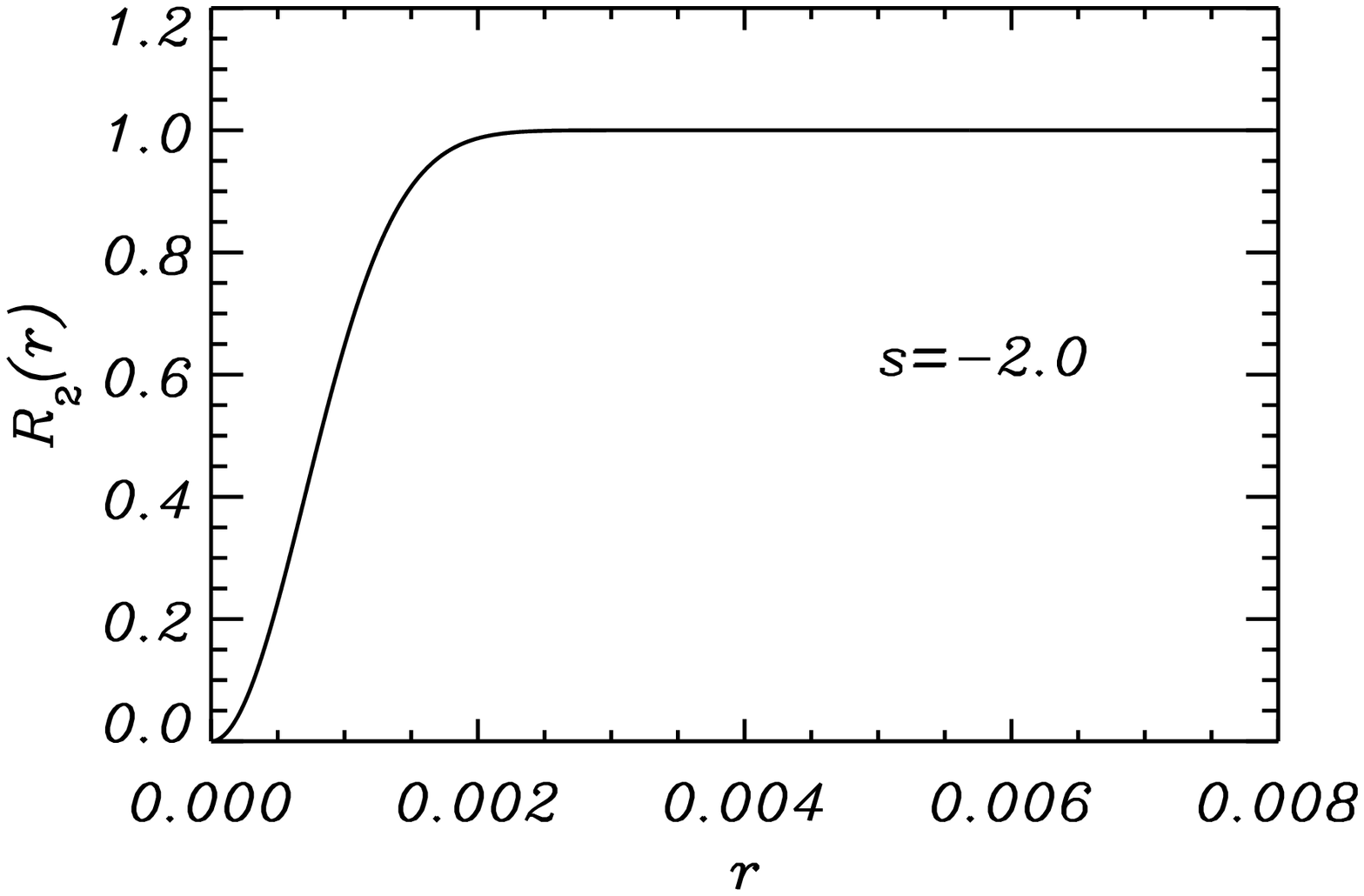}
  \caption{Two--point
  correlation function $R_2(r)$ for different values of
  $s=2\alpha\lambda(\varepsilon-\varepsilon_-)$. For comparison the
  GUE result is shown as well.}
\label{fig2}
\end{figure}

{}From equation (\ref{eq2.41}) the two--point correlation function
results as
\begin{equation}\label{eq4.15}
  R_k(\varepsilon_1,\dots,\varepsilon_k)=(2\alpha)^k
  \mathrm{det}\left[c_{LN}(\varepsilon_p,\varepsilon_q)\right]_{p,q=1,\dots,k}\,,
\end{equation}
where
\begin{equation}\label{4.16}
  c_{LN}(\varepsilon_p,\varepsilon_q)=\frac{f'(z_q)f(z_p)-f(z_q)f'(z_p)}{z_p-z_q}\,.
\end{equation}
(The first factor on the right hand side of equation(\ref{eq4.10})
cancels in taking the determinant as is easily seen.) Just as in
section \ref{sec3.2}, we now introduce rescaled variables
\begin{equation}\label{eq4.17}
  \varepsilon_{pq}=\frac{\varepsilon_p+\varepsilon_q}{2}\ ,\qquad
  Dr_{pq}=\varepsilon_p-\varepsilon_q\,,
\end{equation}
and a rescaled correlation function
$X_k(r_{12},r_{13},\ldots,r_{(k-1)k}) = D^k
R_k(\varepsilon_1,\ldots,\varepsilon_k)$, where
$D^{-1}=\rho(\varepsilon)$ is the local density of states as given by
equation (\ref{eq4.11}). Note that in contrast to section \ref{sec3.2}
we do {\em not} perform the limit $L,N\to\infty$, since we are
interested in particular in the behaviour close to the band edges. We
then have
\begin{equation}\label{eq4.18}
  X_k(r_{12},r_{13},\ldots,r_{(k-1)k})=
  \mathrm{det}
\left[\widehat{c}_{LN}(\varepsilon_{pq},r_{pq})\right]_{p,q=1,\dots,k}
\end{equation}
with
\begin{eqnarray}\label{eq4.19}
 \widehat{c}_{LN}(\varepsilon,r)=c_{LN}\left(\varepsilon+\frac{Dr}{2},
 \varepsilon-\frac{Dr}{2}\right)\,.
\end{eqnarray}
Inserting for $f(z)$ the expression (\ref{eq4.09}), we obtain for the
regime close to the lower edge
\begin{equation}\label{4.19}
 \widehat{c}_{LN}(\varepsilon,r)=\frac
  {\mathrm{Ai}'(-s_+)\mathrm{Ai}(-s_-)-\mathrm{Ai}(-s_+)\mathrm{Ai}'(-s_-)}{r}\,,
\end{equation}
where $s_\pm=s\pm\alpha\lambda Dr$. In the limit $r\to0$ we obtain
$\widehat{c}_{LN}(\varepsilon,0)=1$ as it should be. This is a direct consequence of the
differential equation $\mathrm{Ai}''(z)-z\mathrm{Ai}(z)=0$ of the Airy function, and
equation (\ref{eq4.14}). For $r\to\infty$ $\widehat{c}_{LN}(\varepsilon,r)$ decays
according to

\begin{equation}\label{eq4.19a}
  \widehat{c}_{LN}(\varepsilon,r)\sim r^{-1/4}\exp\left[-\frac{2}{3}
  \left(\alpha\lambda Dr\right)^{2/3}\right]\,,
\end{equation}
which follows from the asymptotic behaviour of the Airy function (see
also equation (\ref{eq4.07})). The transition between the two regimes
is observed at $r=s/\alpha\lambda D$. With decreasing density of
states $\rho(\varepsilon)=1/D$ the transition point is thus
approaching $r=0$, i.\,e. the eigenvalues become more and more
uncorrelated.  This is illustrated in figure \ref{fig2} where the
two--point correlation function
$R_2(r)=1-[\widehat{c}_{LN}(\varepsilon,r)]^2$ is shown for different
values of $s=2\alpha\lambda(\varepsilon-\varepsilon_-)$ in the
neighbourhood of the lower band edge.  In addition the GUE result is
shown for comparison.  We observe with decreasing $s$ a gradual
transition from a GUE behaviour to that expected for completely
uncorrelated eigenvalues.
This is exactly which is expected for a mobility edge: within the band
the eigenvalues experience a quadratic level repulsion typically for
the GUE, whereas in the tails the localization of the wave function
leads to a suppression of the level repulsion (see e.\,g. reference
\cite{Izr}).

\section{Summary and Conclusion}
\label{sec5}

In Ref.~I the density of states for the billiard with randomly
distributed scatterers was calculated, doubly averaged over disorder
and shape of the billiard. We mention in passing that the resulting
model shows some formal similarities to chiral random matrix
models~\cite{SV}. This is due to the way how the average over the
disorder is done. In the present work, the results of Ref.~I are
extended. We calculate the $k$--point correlation functions
exactly. The model of Ref.~I generalizes that of Bogomolny et
al.~\cite{Bogo}. These authors considered a single scatterer in a
chaotic billiard and showed that the fluctuations are chaotic.  In the
present contribution, we extend this study to arbitrarily many
scatterers and also develop a completely different technique to derive
the correlations.  Generalizing the approach of Ref.~\cite{Gu1}, the
correlation functions are expressed in terms of a determinant. This
determinant structure of the correlation functions is immediatly
obvious in the supersymmetric formulation of the model due to the form
of the Berezinian in eigenvalue angle coordinates. Moreover, an
explicit Christoffel--Darboux formula is given for the kernel entering
the determinant.

By means of a saddlepoint appproximation, we rederive the density of
states in the strong coupling limit and also find the real part of the
one--point function. We show that the correlation functions in the
bulk of the spectrum on the scale of the local mean level spacing are,
for all couplings, of GUE type.

Applying a WKB approximation to the kernel, the correlation functions
are studied close to the band edges in the strong coupling limit,
where the number of scatterers is large and the scattering potential
is strong. The above mentioned saddlepoint approximation is not
valid in this regime. Within the band the two--point correlation
function shows a GUE behaviour, but approaching the band edges and
proceeding towards the band tails the eigenvalues become more and more
uncorrelated. This is exactly the fingerprint expected for a mobility
edge and a localization--delocalization transition. We notice that a
drastic depletion of the density of states this accompanies this
transition. Thus, the localization--delocalization transition found in
the non--linear $\sigma$--model~\cite{efe83} is of a different
nature. In the latter, the average is over an ensemble of white--noise
correlated impurities, while two averages are performed in the present
model, one over the wavefunctions at the positions of the scatterers
and another one over the billiard spectrum.  The resulting models are
therefore different. There is a kinetic term in the non--linear
$\sigma$--model and a diffusion constant in front of it. No analogy to
this is present in the model discussed here, because the average over
the billiard spectrum takes care of the kinetic term.

\section*{Acknowledgments}

Part of this work was done while the authors were visiting the Centro
de Ciencias Fisicas, University of Mexico (UNAM), Cuernavaca,
Mexico. We thank Thomas Seligman and the centre for their
hospitality. The microwave experiments motivating this work were
supported by the Deutsche Forschungsgemeinschaft. TG acknowledges
support from Det Svenska Vetenskapsr\aa det.

\appendix

\section{Properties of the Functions Generalizing the
         Hermite Polynomials}
\label{app1}

It is useful to define the functions
\begin{eqnarray}
\varphi_{NL}(\varepsilon) &=& \frac{\exp(-\varepsilon^2/2)}{\sqrt{2^NN!\sqrt{\pi}}}
                              u_{NL}(\varepsilon)
                                   \nonumber\\
\psi_{NL}(\varepsilon) &=& \frac{\exp(-\varepsilon^2/2)}{\sqrt{2^NN!\sqrt{\pi}}}
                              v_{NL}(\varepsilon) \ ,
\label{eqa1.1}
\end{eqnarray}
which reduce to the oscillator wave functions for $L=0$ or,
equivalently, for $\alpha\to\infty$. We also introduce the operators
\begin{eqnarray}
A^+ = \frac{d}{d\varepsilon} - \varepsilon
\qquad {\rm and} \qquad
A^- = \frac{d}{d\varepsilon} + \varepsilon \ ,
\label{eqa1.2}
\end{eqnarray}
which act on the functions~(\ref{eqa1.1}) according to
\begin{eqnarray}
A^+\varphi_{NL}(\varepsilon) &=&
                -\sqrt{2(N+1)}\varphi_{(N+1)L}(\varepsilon)
                                    \nonumber\\
A^-\varphi_{NL}(\varepsilon) &=&
                +\sqrt{2N}\varphi_{(N-1)L}(\varepsilon)
                +\frac{L}{\alpha}\varphi_{N(L+1)}(\varepsilon)
                                    \nonumber\\
A^+\psi_{NL}(\varepsilon) &=&
                -\sqrt{2(N+1)}\psi_{(N+1)L}(\varepsilon)
                -\frac{L}{\alpha}\psi_{N(L-1)}(\varepsilon)
                                    \nonumber\\
A^-\psi_{NL}(\varepsilon) &=&
                +\sqrt{2N}\psi_{(N-1)L}(\varepsilon) \ .
\label{eqa1.3}
\end{eqnarray}
These results extend the formulae for the oscillator wave functions by
terms involving a change of the index $L$. We evaluate the action of
the iterated operators $A^-A^+$ and $A^+A^-$ using
Eqs.~(\ref{eqa1.3}), properly combine terms and arrive at the second
order differential equations
\begin{eqnarray}
\left(\frac{d^2}{d\varepsilon^2}-\varepsilon^2 +(2N+1)\right)
            \varphi_{NL}(\varepsilon) =
-\frac{L}{\alpha}\sqrt{2(N+1)}\varphi_{(N+1)(L+1)}(\varepsilon)
                                    \nonumber\\
\left(\frac{d^2}{d\varepsilon^2}-\varepsilon^2 +(2N+1)\right)
            \psi_{NL}(\varepsilon) =
-\frac{L}{\alpha}\sqrt{2N}\psi_{(N-1)(L-1)}(\varepsilon)
\label{eqa1.4}
\end{eqnarray}
These are no eigenvalue equations, because the functions on the left
and the right hand sides have different indices. However, one can cast
them into diffusion--type--of equations by introducing the fictitious
time
\begin{eqnarray}
\tau = -\ln\alpha
\qquad {\rm such \ that} \qquad
\alpha = \exp(-\tau) \ .
\label{eqa1.5}
\end{eqnarray}
A straightforward calculation yields the equations
\begin{eqnarray}
\left(\frac{d^2}{d\varepsilon^2}-\varepsilon^2 +(2N+1)\right)
            \varphi_{NL}(\varepsilon) =
-2\frac{\partial}{\partial\tau}\varphi_{NL}(\varepsilon)
                                    \nonumber\\
\left(\frac{d^2}{d\varepsilon^2}-\varepsilon^2 +(2N+1)\right)
            \psi_{NL}(\varepsilon) =
+2\frac{\partial}{\partial\tau}\psi_{NL}(\varepsilon) \ ,
\label{eqa1.6}
\end{eqnarray}
which involve the same indices on both sides.


\end{document}